\documentclass[paper a4]{article}
\title{Noncommutative relativistic particle on the electromagnetic
background.}
\author{A.A. Deriglazov\footnote{alexei@fisica.ufjf.br ~ On leave of
absence from Dept. Math. Phys., Tomsk Polytechnical University,
Tomsk, Russia.}}
\date{Dept. de Matematica, ICE, Universidade Federal de Juiz de Fora,\\
MG, Brasil.}
\begin{document}
\maketitle
\large
\begin{abstract}
Noncommutative version of D-dimensional relativistic particle is proposed.
We consider the particle interacting with the configuration space
variable $\theta^{\mu\nu}(\tau)$ instead of the numerical matrix. The
corresponding Poincare invariant action has a local symmetry, which
allows one to impose the gauge $\theta^{0i}=0, ~ \theta^{ij}=const$. The
matrix $\theta^{ij}$ turns out to be the noncommutativity parameter of
the gauge fixed formulation. Poincare transformations of the gauge
fixed formulation are presented in the manifest form.
Consistent coupling of NC relativistic particle to the electromagnetic
field is constructed and discussed.
\end{abstract}

{\bf PAC codes:} 0460D, 1130C, 1125 

\noindent
\section{Introduction.}
It is known that the
noncommutative (NC) geometry [1, 2] of the position variables in
some mechanical
models can be obtained [3-7] as the result of direct canonical quantization
[9, 10] of underlying dynamical
systems with second class constraints. Nontrivial bracket for
the position variables appears in this case as the Dirac bracket,
after taking into account the constraints presented in the model.
An apparent defect of the known NC models is lack of relativistic invariance,
due to the fact that noncommutativity parameter is constant matrix.

In this note we discuss one possibility to resolve the problem. Namely,
the noncommutative version for D-dimensional
relativistic particle is proposed. We show also that it is possible to write
(rather exotic) interaction with an external electromagnetic field.
The interaction is consistent with the Poincare invariance, as well as
with local symmetries presented in the model.

The work is organized as follows. In Sec. 2 we demonstrate that a
procedure used to obtain NC versions of the particular models
[3, 6, 7] can be generalized to the case of an arbitrary nondegenerate
mechanical system. Namely, to obtain NC version, it is
sufficiently to add Chern-Simons type term to the first order Lagrangian
action of the initial system. The numerical matrix
$\theta^{AB}=-\theta^{BA}$, originating from the Chern-Simons term,
turns out to be the NC parameter of the formulation.
We point also that quantization of the NC system leads to quantum
mechanics with ordinary product replaced by the Moyal product.

In Sec. 3 we show that a slight modification of the procedure allows one
to obtain NC version for $D$-dimensional relativistic
particle. Chern-Simons term can
be added to the first order action of the relativistic particle, which do
not spoil the reparametrization invariance. As a consequence,
the model will contain the desired relativistic constraint $p^2-m^2=0$.
The problem is
that the numerical matrix $\theta^{\mu\nu}$ do not respect the Lorentz
invariance. To resolve the problem, we consider a particle interacting
with a new configuration-space variable
$\theta^{\mu\nu}(\tau)=-\theta^{\nu\mu}(\tau)$,
instead of the constant matrix. The action constructed is manifestly
Poincare invariant and has local symmetry related with the variable
$\theta$. The last one can be gauged out, an admissible gauge is
$\theta^{0i}=0, ~ \theta^{ij}=const$. The noncommutativity parameter of the
gauge fixed version is then the numerical matrix $\theta^{ij}$. As it 
usually happens in a
theory with local symmetries [11], Poincare invariance of the gauge
fixed version is combination of the initial Poincare and local
transformations which preserve the gauge chosen. In the case under
consideration, the resulting transformations are linear and
involve the constant matrix $\theta^{ij}$ (see Eqs.(\ref{29}) below).

In Sec. 4 interaction with an external electromagnetic field is discussed.
The standard interaction term can be added, in principle, but does not
lead to an interesting situation. Consistency of the term with the
local symmetries presented in the model implies it's specific dependence
on the configuration space variables. As a consequence, after transition
to the canonical variables,
any traces of noncommutativity disappear from the formulation.
As an alternative, we propose new interaction term which involve the
field strength instead of the electromagnetic potential. The possibility to
write the term is implied by the fact that one works with the first
order Lagrangian action \footnote{Let us note that the same
interaction term can be written for ordinary relativistic particle
(in the first order formulation) as well.}.
In the Conclusion we discuss combined interaction.

\section{Noncommutative version of an arbitrary nondegenerate mechanics.}

Our starting point is some nondegenerate
mechanical system with the configuration space variables $q^A(t), ~
A=1,2, \ldots , n$, and the Lagrangian action
\begin{eqnarray}\label{1}
S=\int dt L(q^A, ~ \dot q^A).
\end{eqnarray}
Due to nondegenerate character of the system, there are no constraints
in the Hamiltonian formulation. Let $p_A$ are conjugated momentum for
$q^A$, one can write the Hamiltonian action
\begin{eqnarray}\label{2}
S_H=\int dt \left[ p_A\dot q^A-H_0(q^A, ~ p_A)\right].
\end{eqnarray}
Equations of motion which follow from Eq.(\ref{1}) and (\ref{2}) are
equivalent (they remain equivalent for any degenerated system also
[10, 12], in this case the Hamiltonian includes the Lagrangian
multipliers). Equivalently,
one can describe the initial system (\ref{1}) by means of the first order
{\em Lagrangian action}
\begin{eqnarray}\label{3}
S_1=\int dt \left[v_A\dot q^A-H_0(q^A, ~ v_A)\right].
\end{eqnarray}
Here $q^A(t), ~ v_A(t)$ are the configuration space variables of the
formulation \footnote{The Lagrangian formulations (\ref{1}), (\ref{3})
are equivalent. Actually, denoting the conjugated momentum for the
variables $q^A, ~ v_A$ as $p_A, ~ \pi^A$ one finds, in the Hamiltonian
formulation for the action (\ref{3}), the second class constraints
$p_A-v_A=0, ~ \pi^A=0$. Introducing the corresponding Dirac bracket,
one can treat the constraints as the strong equations. Then the
Hamiltonian formulation for (\ref{3}) is the same as for (\ref{1}),
namely Eq.(\ref{2}).}.
The noncommutative version of the system (\ref{1}) is described by the
following Lagrangian action
\begin{eqnarray}\label{4}
S_N=\int dt \left[v_A\dot q^A-H_0(q^A, ~ v_A)+
\dot v_A\theta^{AB}v_B \right],
\end{eqnarray}
where $\theta^{AB}$ is some constant matrix. It turns out to be the
noncommutativity parameter for the variables $q^A$.

Let us analyse the model (\ref{4}) in the Hamiltonian framework (see
[8] for details). All the expressions for determining
of the momentum turn out to be the primary constraints of the model ~
($p_A,  \pi^A$ are conjugated momentum for the variables $q^A,  v_A$)
\begin{eqnarray}\label{5}
G_A\equiv p_A-v_A=0, \qquad
T^A\equiv\pi^A-\theta^{AB}v_B=0,
\end{eqnarray}
with the Poisson bracket algebra being of second class
\begin{eqnarray}\label{6}
\{G_A, G_B\}=0, \qquad \{T^A, T^B\}=
-2\theta^{AB}, \qquad
\{G_A, T^B\}=-\delta_A^B.
\end{eqnarray}
The constraints can be taken into account by transition to the Dirac
bracket. After that, one can take the variables $(q^A, ~ p_A)$ as the
physical one, while $(v_A, ~ \pi^A)$ can be omitted from consideration
using Eq.(\ref{5}). The resulting noncommutative system has the
following properties. \\
1) It has the same number of physical degrees of freedom as the initial
system $S$, namely $q^A, ~ p_A$. \\
2) Equations of motion of the system are the same as for the initial
system $S$, modulo the term which is proportional to the parameter
$\theta^{AB}$
\begin{eqnarray}\label{7}
\dot q^A=\frac{\partial H_0}{\partial p_A}-2\theta^{AB}
\frac{\partial H_0}{\partial q^B}, \qquad
\qquad \dot p_A=-\frac{\partial H_0}{\partial q^A},
\end{eqnarray}
where $H_0(q, ~ p)=H_0(q, ~ v)|_{v\rightarrow p}$. \\
3) The physical variables have the brackets
\begin{eqnarray}\label{8}
\{q^A, q^B\}=-2\theta^{AB}, \qquad
\{q^A, p_B\}=\delta^A_B,
\qquad \{p_A, p_B\}=0.
\end{eqnarray}
In particular, brackets of the configuration space variables are
noncommutative. One can show that other possibilities to choose the
physical variables: ~ $q^A, ~ v_A$, or $q^A, ~ \pi_A$ ~ lead to an
equivalent description.

To quantize the resulting system, one possibility is to find variables
which have the canonical brackets. For the case under consideration
they are
\begin{eqnarray}\label{9}
\tilde q^A=q^A-\theta^{AB}p_B, \qquad \tilde p_A=p_A,
\end{eqnarray}
and obey $\{\tilde q, \tilde q\}=\{\tilde p, \tilde p\}=0, ~
\{\tilde q, \tilde p\}=1$. Equations of motion in terms
of these variables acquire the standard form
\begin{eqnarray}\label{10}
\dot{\tilde q}^A=\{\tilde q^A, \tilde H_0\}, \qquad
\dot{\tilde p}_A=\{\tilde p_A, \tilde H_0\},
\end{eqnarray}
where $\tilde H_0=H_0(\tilde q+\theta\tilde p, ~ \tilde p)$. It leads to
quantum
mechanics with the Moyal product (see [7] and references therein)
\begin{eqnarray}\label{11}
H_0(\tilde q^A+\theta^{AB}\tilde p_B, ~ \tilde p_B)\Psi(\tilde q^C)=
H_0(\tilde q^A, ~ \tilde p_B)*\Psi(\tilde q^C).
\end{eqnarray}
The procedure described above can be
applied to some degenerated systems as well. The necessary condition
is that a part of
variables enter into the initial action without the time derivatives,
and such that they can be identified with
the Lagrangian multipliers of the Hamiltonian formulation.
Then the system admits the first order {\em Lagrangian formulation}
(\ref{3}). The relativistic
particle and the string are examples of such a system (see [13] for the
first order formulation of the string). We suppose that the procedure
can be applied to the spinning particle [14] and to the superparticle [15].
It may be interesting [16] since both models are supersymmetric.
If the relativistic invariance is
presented in the initial formulation, a slight modification of
the procedure is required to keep the symmetry in the NC version. The
modification is presented in the next section.

\section{Noncommutative relativistic particle.}

The configuration space variables of the model
are $x^\mu(\tau), ~ v^\mu(\tau), ~ e(\tau)$, $\theta^{\mu\nu}(\tau)$,
with the Lagrangian action being
\begin{eqnarray}\label{12}
S=\int d\tau\left[\dot x^\mu v_\mu-\frac{e}{2}(v^2-m^2)+
\frac{1}{\theta^2}\dot v_\mu\theta^{\mu\nu}v_\nu\right].
\end{eqnarray}
Here $\theta^2\equiv\theta^{\mu\nu}\theta_{\mu\nu}, ~ \eta^{\mu\nu}=
(+,-, \ldots ,-)$. Insertion of the term $\theta^2$
in the denominator has the same meaning as for the eibein in
the action of massless particle: $L=\frac{1}{2e}\dot x^2$. Technically,
it rules out the degenerated gauge $e=0$. The action is manifestly
invariant under the Poincare transformations
\begin{eqnarray}\label{13}
x'^\mu=\Lambda^\mu{}_\nu x^\nu+a^\mu, \quad v'^\mu=\Lambda^\mu{}_\nu
v^\nu, \quad
\theta '^{\mu\nu}=\Lambda^\mu{}_\rho\Lambda^\nu{}_\sigma
\theta^{\rho\sigma}.
\end{eqnarray}
Local symmetries of the model are reparametrizations (with
$\theta^{\mu\nu}$ being the scalar variable), and the following
transformations with the parameter $\epsilon_{\mu\nu}(\tau)=-
\epsilon_{\nu\mu}(\tau)$
\begin{eqnarray}\label{14}
\delta x^\mu=-\epsilon^{\mu\nu}v_\nu, \qquad
\delta\theta_{\mu\nu}=-\theta^2\epsilon_{\mu\nu}+
2\theta_{\mu\nu}(\theta\epsilon).
\end{eqnarray}
To analyse physical sector of this constrained system, we rewrite
it in the Hamiltonian form.
Starting from the action (\ref{12}), one finds in the Hamiltonian
formalism the primary constraints
\begin{eqnarray}\label{15}
G^\mu\equiv p^\mu-v^\mu=0, \qquad
T^\mu\equiv\pi^\mu-\frac{1}{\theta^2}\theta^{\mu\nu}v_\nu=0, \cr
p_{\theta}^{\mu\nu}=0, \qquad \qquad p_e=0,
\end{eqnarray}
and the Hamiltonian
\begin{eqnarray}\label{16}
H=\frac{e}{2}(v^2-m^2)+\lambda_{1\mu}G^\mu+\lambda_{2\mu}T^\mu+
\lambda_ep_e+\lambda_{\theta\mu\nu}p_{\theta}^{\mu\nu}.
\end{eqnarray}
Here $p, ~ \pi$ are conjugated momentum for $x, ~ v$ and $\lambda$ are the
Lagrangian multipliers for the constraints. On the next step there is
appear the secondary constraint
\begin{eqnarray}\label{17}
v^2-m^2=0,
\end{eqnarray}
as well as equations for determining the Lagrangian multipliers
\begin{eqnarray}\label{18}
\lambda_2^\mu=0, \qquad
\lambda_1^\mu=ev^\mu+\frac{2}{\theta^2}(\lambda_\theta v)^\mu-
\frac{4}{\theta^4}(\theta\lambda_{\theta})(\theta v)^\mu.
\end{eqnarray}
There is no of tertiary constraints in the problem.
Equations of motion follow from (\ref{16})-(\ref{18}), in particular,
for the variables $x, ~ p$ one has
\begin{eqnarray}\label{19}
\dot x^\mu=ep^\mu+\frac{2}{\theta^2}(\lambda_\theta v)^\mu-
\frac{4}{\theta^4}(\theta\lambda_{\theta})(\theta v)^\mu, \qquad
\qquad \dot p^\mu=0
\end{eqnarray}
Poisson brackets of the constraints are
\begin{eqnarray}\label{20}
\{G^\mu, G^\nu\}=0, \qquad \qquad \{T^\mu, T^\nu\}=
-\frac{2}{\theta^2}\theta^{\mu\nu}, \cr
\{G_\mu, T^\nu\}=-\delta_\mu^\nu, \qquad
\{T_\mu, p_\theta^{\rho\sigma}\}=-\frac{1}{\theta^2}
\delta_\mu^{[\rho}v^{\sigma ]}+
\frac{4}{\theta^4}(\theta v)_\mu\theta^{\rho\sigma}.
\end{eqnarray}
The constraints $G^\mu, ~ T^\mu$ form the second class subsystem and can
be taken into account by transition to the Dirac bracket. Then the
remaining constraints can be classified in accordance with their properties
relatively to the Dirac bracket. Consistency of the procedure is
guaranteed by the known theorems [10]. Introducing the Dirac bracket
\begin{eqnarray}\label{21}
\{A, B\}_D=\{A, B\}+\{A, G_\mu\}\frac{2}{\theta^2}\theta^{\mu\nu}
\{G_\nu, B\}- \cr
\{A, G^\mu\}\{T_\mu, B\}+\{A, T_\mu\}\{G^\mu, B\},
\end{eqnarray}
one finds, in particular, the following brackets for the
fundamental variables (all the nonzero brackets are presented)
\begin{eqnarray}\label{22}
\{x^\mu, x^\nu\}=-\frac{2}{\theta^2}\theta^{\mu\nu}, \quad
\{x^\mu, p_\nu\}=\delta^\mu_\nu,
\quad \{p_\mu, p_\nu\}=0;
\end{eqnarray}
\begin{eqnarray}\label{23}
\{x^\mu, v_\nu\}=\delta^\mu_\nu, \quad
\{x^\mu, \pi^\nu\}=-\frac{1}{\theta^2}\theta^{\mu\nu}, \quad
\{\theta_{\mu\nu}, p_\theta^{\rho\sigma}\}=-\delta_\mu^{[\rho}
\delta_\nu^{\sigma ]}, \cr
\{x^\mu, p_\theta^{\rho\sigma}\}=-\{\pi^\mu, p_\theta^{\rho\sigma}\}=
\frac{1}{\theta^2}\eta^{\mu [\rho}v^{\sigma ]}-
\frac{4}{\theta^4}(\theta v)^\mu\theta^{\rho\sigma}.
\end{eqnarray}
Let us choose $x^\mu, ~ p^\mu$ as the physical sector variables (one can
equivalently take $(x, v)$ or $(x, \pi)$, which leads to the same final
results, similarly to the non relativistic case [7]).
The variables $v, \pi$ can be omitted now from the consideration.

Up to now the procedure preserves the manifest Poincare invariance of
the model. Let us discuss the first class constraints
$p_\theta^{\rho\sigma}=0$. As the gauge fixing conditions one takes
\begin{eqnarray}\label{24}
\theta^{0i}=0, \qquad \qquad \theta^{ij}=const.
\end{eqnarray}
Then $\theta^{\mu\nu}\theta_{\mu\nu}=
-\theta_{ij}\theta_{ji}$, and the gauge is admissible if
$\theta_{ij}\theta_{ji}\ne 0$, see Eq.(\ref{22}), (\ref{19}). From the
equation of motion
$\dot\theta=\lambda_\theta$ one determines the remaining Lagrangian
multipliers: $\lambda_\theta=0$. Using this result in Eq.(\ref{19}),
the final form of the equations of motion is
\begin{eqnarray}\label{25}
\dot x^\mu=ep^\mu, \qquad \dot p^\mu=0.
\end{eqnarray}
They are supplemented by the
remaining first class constraints $p^2-m^2=0, ~ p_e=0$. Brackets for
the physical variables are given by Eqs.(\ref{22}).

The initial Poincare transformations (\ref{13}) do not preserve the gauge
(\ref{24}) and must be accompanied by compensating local transformation,
with the parameter $\epsilon^{\mu\nu}$ chosen in appropriate way.
It gives the Poincare symmetry of the gauge fixed version. To find it,
one has the conditions ($\Lambda^\mu{}_\nu=\delta^\mu_\nu+
\omega^\mu{}_\nu$)
\begin{eqnarray}\label{26}
(\delta_\omega+\delta_\epsilon)\theta^{0i}=\omega^0{}_j\theta^{ji}+
\theta^2\epsilon^{0i}=0, \cr
(\delta_\omega+\delta_\epsilon)\theta^{ij}=\omega^{[i}{}_k\theta^{kj]}
\theta^2\epsilon^{ij}+2\theta^{ij}(\theta^{kp}\epsilon_{kp})=0
\end{eqnarray}
The solution is
\begin{eqnarray}\label{27}
\epsilon^{0i}(\omega)=\frac{1}{\theta^2}\omega^{0j}\theta^{ji}, \quad
\epsilon^{ij}(\omega)=\frac{1}{\theta^2}\omega^{[i}{}_k\theta_k{}^{j]},
\end{eqnarray}
or, equivalently
\begin{eqnarray}\label{28}
\epsilon^{\mu\nu}(\omega)=-\frac{1}{\theta^2}
\omega^{[\mu}{}_\rho\theta^{\rho\nu]},
\end{eqnarray}
where Eq.(\ref{24}) is implied. Then the Poincare transformations of
the gauge fixed version are
\begin{eqnarray}\label{29}
\delta x^\mu=\omega^\mu{}_\nu x^\nu+\frac{1}{\theta^2}p_\nu
\omega^{[\nu}{}_\rho\theta^{\rho\mu]}, \quad
\delta p^\mu=\omega^\mu{}_\nu p^\nu.
\end{eqnarray}

\section{Interaction with an external electromagnetic field.}

The standard interaction term $A_\mu(x)\dot x^\mu$ can not be added to the
NC action (\ref{12}), since it will break the local symmetry (\ref{14}).
To preserve the symmetry, one needs to take the electromagnetic field
depending on the gauge-invariant combination:
$A_\mu\left(x-\frac{\theta v}{\theta^2}\right)\dot x^\mu$. Then, in
terms of the
canonical variables (\ref{9}), any traces of noncommutativity disappear
in the final formulation, similarly to the free particle case.

Other natural possibility, which is implied by the first order formulation,
is coupling to the field strenght of the form
$\dot v_\mu F^{\mu\nu}v_\nu$, where
$F_{\mu\nu}=\partial_{[\mu}A_{\nu]}(x)$. So, let us consider the action
\begin{eqnarray}\label{30}
S=\int d\tau\left[\dot x^\mu v_\mu-\frac{e}{2}(v^2-m^2)+
\frac{1}{\theta^2}\dot v_\mu\theta^{\mu\nu}v_\nu+
\dot v_\mu F^{\mu\nu}v_\nu\right].
\end{eqnarray}
Note that the interaction term can not be removed by shift of the
$\theta$-variable, due to presence of $\theta^2$ in the denominator. Local
symmetries of the model are reparametrizations, $U(1)$ gauge
transformations, and the modified $\epsilon$ transformations which look
now as follow:
\begin{eqnarray}\label{31}
\delta x^\mu=-\epsilon^{\mu\nu}v_\nu, \qquad
\delta\theta_{\mu\nu}=-\theta^2(\epsilon_{\mu\nu}+
\delta_\epsilon F_{\mu\nu})+ \cr 
2\theta_{\mu\nu}\left[(\theta\epsilon)+(\theta\delta_\epsilon F)\right].
\end{eqnarray}
Hamiltonian analysis of the model is similar to the free particle case
discussed above. The interaction term leads to deformation of the
constraint structure as compare with (\ref{15})-(\ref{23}). The
constraints of the model are
\begin{eqnarray}\label{32}
G^\mu\equiv p^\mu-v^\mu=0, \qquad
T^\mu\equiv\pi^\mu-\frac{1}{\theta^2}\theta^{\mu\nu}v_\nu-
F^{\mu\nu}v_\nu=0, \cr
p_{\theta}^{\mu\nu}=0, \qquad p_e=0, \qquad
v^2-m^2=0.
\end{eqnarray}
The Poisson bracket algebra is deformed also and acquires the form
\begin{eqnarray}\label{33}
\{G^\mu, G^\nu\}=0, \qquad \qquad \{T^\mu, T^\nu\}=
\Delta^{\mu\nu}, \cr
\{T^\mu, G^\nu\}=\Delta_1{}^{\mu\nu}, \qquad
\{G^\mu, T^\nu\}=-(\Delta_1^T)^{\mu\nu},
\end{eqnarray}
where it was denoted
\begin{eqnarray}\label{34}
\Delta^{\mu\nu}=-2\left(\frac{\theta^{\mu\nu}}{\theta^2}+
F^{\mu\nu}\right), \qquad
\Delta_1{}^{\mu\nu}=\eta^{\mu\nu}-\partial^\nu F^{\mu\rho}v_\rho.
\end{eqnarray}
The only nonzero bracket of the constraint
$p_\theta^{\mu\nu}=0$ with others is the same as in (\ref{20}).
The Dirac bracket which corresponds to the constraints $G^\mu=0, ~
T^\mu=0$ is
\begin{eqnarray}\label{35}
\{A, B\}_D=\{A, B\}-\{A, G_\mu\}\left(\Delta_1^{-1}\Delta
\Delta_1^{-1{}T}\right)^{\mu\nu}\{G_\nu, B\}- \cr
\{A, G_\mu\}(\Delta_1^{-1})^{\mu\nu}\{T_\nu, B\}+\{A, T_\mu\}
(\Delta_1^{-1{}T})^{\mu\nu}\{G_\nu, B\}.
\end{eqnarray}
One takes the same gauge as in (\ref{24}) for the  first class
constraints $p_\theta^{\mu\nu}=0$, and the gauge $e=1$ for $p_e=0$.
The resulting system can be taken into account by transition to
the corresponding Dirac bracket. Then the remaining variables of the
theory are ~ $x^\mu, p_\mu$. Classical dynamics of these variables is
described by the Hamiltonian equations
\begin{eqnarray}\label{36}
\dot x^\mu=(\Delta_1^{-1})^{\mu\nu}p_\nu, \qquad \dot p^\mu=0,
\end{eqnarray}
which are accompanied by the constraint $p^2-m^2=0$. Brackets for the
variables are
\begin{eqnarray}\label{37}
\{x^\mu, x^\nu\}=\left(\Delta_1^{-1}\Delta
\Delta_1^{-1{}T}\right)^{\mu\nu}, \cr 
\{x^\mu, p_\nu\}=\delta^\mu{}_\nu+\left(\Delta_1^{-1}\right)^{\mu\rho}
\partial_\nu F^{\rho\sigma}p_\sigma, \qquad
\{p_\mu, p_\nu\}=0.
\end{eqnarray}
One notes that in terms of the variables $x, p$ the
noncommutativity parameter $\theta$ does not enter into the equations of
motion. From Eq.(\ref{36}) one finds the second order equation for the
position variable
\begin{eqnarray}\label{38}
\ddot x^\mu=(\Delta_1^{-1})^{\mu\sigma}\partial_\alpha\partial_\beta
F_{\sigma\rho}\Delta_1{}^{\rho\nu}\dot x^\nu\dot x^\alpha\dot x^\beta.
\end{eqnarray}
Interesting property of the interaction is that dynamics of NC variables
in the constant electromagnetic field is governed by free equation. All the
information on dynamics is encoded in this case in the noncommutative
brackets (\ref{37}). For the non relativistic systems the same property
was discussed in [7]. Let us point also that it may be mechanical analogy
of duality relations [17].

\section{Conclusion.}

In this work we have presented noncommutative version (ref{12})of
$D$-dimensional relativistic particle. It couples to the electromagnetic
background through the field strength, see (\ref{30}). The interaction
introduced is consistent with the Poincare invariance as well as with
local symmetries presented in the model. Some relevant
comments are in order. \\
1) The same interaction term can be added to the first order action of
usual (commutative) relativistic particle as well. It may be interesting
to study this trick in the context of higher spin particle models [18]
(it is well known that the standard coupling is not consistent with
symmetries of higher spin actions [19-21]). \\
2) Let us point that in the second order formulation, a similar
interaction term could be $F_{\mu\nu}\dot x^\mu\ddot x^\nu$. One expects
that it will lead to different physical picture as compare with 1).
The term involve the higher derivative, which indicates on appearance
of extra physical degrees of freedom. \\
3) At last, we point that the standard coupling can be combined with
the one considered in Sec. 4, one takes
\begin{eqnarray}\label{39}
S_{int}=\int d\tau\left[A_\mu\left(x-\frac{\theta v}{\theta^2}\right)
\dot x^\mu+\dot v_\mu F^{\mu\nu}(x)v_\nu\right].
\end{eqnarray}
Since the bracket algebra (\ref{37}) is deformed
as compare with the free case (due to presence of
the field strength term), this interacting
system will be different from the corresponding commutative one.

\end{document}